\documentclass[aps,preprint]{revtex4}%
\usepackage{amsfonts}
\usepackage{amsmath}
\usepackage{amssymb}
\usepackage{graphicx}%
\begin{document}
\title[ ]{ Suppression of Weiss oscillations in the magnetoconductivity of modulated
graphene monolayer}
\preprint{ }
\author{M. Tahir$^{\ast}$}
\affiliation{Department of Physics, University of Sargodha, Sargodha 40100, Pakistan}
\author{K. Sabeeh$^{\dagger}$}
\affiliation{Department of Physics, Quaid-i-Azam University, Islamabad 45320, Pakistan}
\author{}
\affiliation{}
\pacs{PACS number}

\begin{abstract}
We have investigated the electrical transport properties of Dirac electrons in
a monolayer graphene sheet in the presence of both electric and magnetic
modulations. The effects of the modulations on quantum transport when they are
in phase and out of phase are considered. We present the energy spectrum and
the bandwidth of the Dirac electrons in the presence of both the modulations.
We determine the $\sigma_{yy}$ component of the magnetoconductivity tensor for
this system which is shown to exhibit Weiss oscillations.Asymptotic
expressions for $\sigma_{yy}$ are also calculated to better illustrate the
effects of in-phase and out-of-phase modulations.We find that the position of
the oscillations in magnetoconductivity depends on the relative strength of
the two modulations. When the two modulations are out-of-phase there is
complete suppression of Weiss oscillations for particular relative strength of
the modulations.

\end{abstract}
\volumeyear{year}
\volumenumber{number}
\issuenumber{number}
\eid{identifier}
\date[Date text]{date}
\received[Received text]{date}

\revised[Revised text]{date}

\accepted[Accepted text]{date}

\published[Published text]{date}

\startpage{1}
\endpage{2}
\maketitle

\section{\textbf{Introduction }}

The recent successful preparation of monolayer graphene has generated a lot of
interest in the physics community and efforts are underway to study the
electronic properties of graphene \cite{1}. The nature of quasiparticles
called Dirac electrons in these two-dimensional systems is very different from
those of the conventional two-dimensional electron gas (2DEG) realized in
semiconductor heterostructures. Graphene has a honeycomb lattice of carbon
atoms. The quasiparticles in graphene have a band structure in which electron
and hole bands touch at two points in the Brillouin zone. At these Dirac
points the quasiparticles obey the massless Dirac equation. In other words,
they behave as massless Dirac particles leading to a linear dispersion
relation $\epsilon_{k}=vk$ ( with the characteristic velocity $v\simeq
10^{6}m/s)$. This difference in the nature of the quasiparticles in graphene
from conventional 2DEG has given rise to a host of new and unusual phenomena
such as anomalous quantum Hall effects and a $\pi$ Berry phase\cite{1}%
\cite{2}. Among the electronic properties of interest is the interaction of
electrons with artificially created periodic potentials. It has been observed
that if conventional 2DEG is subjected to artificially created periodic
potentials in the submicrometer range it leads to the appearance of Weiss
oscillations in the magnetoresistance. This type of electrical modulation of
the 2D system can be carried out by depositing an array of parallel metallic
strips on the surface or through two interfering laser beams \cite{3,4,5}.
Besides the fundamental interest in understanding the electronic properties of
graphene there is also serious suggestions that it can serve as the building
block for nanoelectronic devices \cite{6}.

In conventional 2DEG systems, electron transport in the presence of electric
and magnetic modulation has continued to be an active area of research
\cite{7,8,9}. Recently, electrical transport in graphene monolayer in the
presence of electrical modulation was considered and theoretical predictions
made \cite{10}. We have also carried out a study of magnetoconductivity when
graphene monolayer is subjected to magnetic modulation alone\cite{11}. Along
the same lines, in this work we investigate low temperature magnetotransport
of Dirac electrons in a graphene monolayer subjected to both electric and
magnetic modulations with the same period. From a practical point of view,
this is important to consider as magnetic modulation of graphene can be
realized by magnetic or superconducting stripes placed on top of graphene
which in turn act as electrical gates that induce electric modulation. The
relative phase of the two modulations can have important consequences for
magnetotransport in the system as was seen in conventional 2DEG. Therefore in
this work we investigate Weiss oscillations in magnetoconductivity
$\sigma_{yy}$ in a graphene monolayer for both the cases when the electric and
magnetic modulations are in-phase and when they are out of phase.

\section{Energy Spectrum and Bandwidth}

We consider two-dimensional Dirac electrons in graphene moving in the
$x-y-$plane. The magnetic filed ($B$) is applied along the $z-$direction
perpendicular to the graphene plane. We consider the perpendicular magnetic
field $B$ modulated weakly and periodically along one direction such that
$\mathbf{B}=(B+B_{0}\cos(Kx))\mathbf{z}$. Here $B_{0}$ is the strength of the
magnetic modulation and $K=2\pi/a$ with $a$ being the modulation period. We
consider the modulation to be weak such that $B_{0}<<B$. We use the Landau
gauge for vector potential $\mathbf{A}=(0,Bx,0)$. In effective mass
approximation the one electron Hamiltonian is $H=v\overleftrightarrow{\sigma
}.(-i\hbar\mathbf{\nabla}+e\mathbf{A}).$\ The low energy excitations are
described by the two-dimensional (2D) Dirac like Hamiltonian \cite{1,2,10}%
$H=v\overleftrightarrow{\sigma}.(-i\hbar\mathbf{\nabla}+e\mathbf{A}%
).$Here\ $\overleftrightarrow{\sigma}=\{\overleftrightarrow{\sigma}%
_{x},\overleftrightarrow{\sigma}_{y}\}$ are the Pauli matrices and $v$
characterizes the electron velocity. We employ the Landau gauge and write the
vector potential as $\mathbf{A}=(0,Bx+(B_{0}/K)\sin(Kx),0)$. The Hamiltonian
can be written as $H=H_{0}+H_{,}^{\prime}$where $H_{0}$ is the unmodulated
Hamiltonian given as $H_{0}=-i\hbar v\overleftrightarrow{\sigma}%
.\mathbf{\nabla}+ev\overleftrightarrow{\sigma_{y}}Bx$ and $H^{\prime
}=ev\overleftrightarrow{\sigma_{y}}\frac{B_{0}}{K}\sin(Kx).$The Landau level
energy eigenvalues without modulation are given by%
\begin{equation}
\varepsilon(n)=\hbar\omega_{g}\sqrt{n} \label{1}%
\end{equation}
where $n$ is an integer and $\omega_{g}=v\sqrt{2eB/\hbar}.$ As has been
pointed out \cite{10} the Landau level spectrum for Dirac electrons is
significantly different from the spectrum for electrons in conventional
2DEG\ which is given as $\varepsilon(n)=\hbar\omega_{c}(n+1/2)$, where
$\omega_{c}=eB/m$ is the cyclotron frequency.

The eigenfunctions without modulation are given by
\begin{equation}
\Psi_{n,k_{y}}(r)=\frac{e^{ik_{y}y}}{\sqrt{2L_{y}}}\left(
\begin{array}
[c]{c}%
-i\varphi_{n-1}(x,x_{0})\\
\varphi_{n}(x,x_{0})
\end{array}
\right)  \label{2}%
\end{equation}
where $\varphi_{n}(x,x_{0})=\frac{e^{-(x+x_{0})/2l}}{\sqrt{2^{n}n!l\sqrt{\pi}%
}}H_{n}(\frac{x+x_{0}}{l})$ are the normalized harmonic oscillator
eigenfunctions$,l=\sqrt{\hbar/eB}$ is the magnetic length, $x_{0}=l^{2}k_{y},$
$L_{y}$ is the $y$-dimension of the graphene layer and $H_{n}(x)$ are the
Hermite polynomials.

Since we are considering weak modulation $B_{0}<<$ $B$, we can apply standard
perturbation theory to determine the first order corrections to the
unmodulated energy eigenvalues in the presence of modulation with the result
$\Delta\varepsilon_{_{n,k_{y}}}=\omega_{0}\cos(Kx_{0})\left(  \sqrt{\frac
{2n}{u}}e^{-u/2}[L_{n-1}(u)-L_{n}(u)]\right)  $where $\omega_{0}=\frac
{evB_{0}}{K}$, $u=K^{2}l^{2}/2$ and $L_{n}(u)$ are the Laguerre polynomials.
Hence the energy eigenvalues in the presence of the periodic magnetic
modulation are
\begin{equation}
\varepsilon(n,x_{0})=\varepsilon(n)+\Delta\varepsilon_{_{n,k_{y}}}=\hbar
\omega_{g}\sqrt{n}+\omega_{0}\cos(Kx_{0})G_{n} \label{3}%
\end{equation}
with\ $G_{n}(u)=\sqrt{\frac{2n}{u}}e^{-u/2}[L_{n-1}(u)-L_{n}(u)]$. We observe
that the degeneracy of the Landau level spectrum of the unmodulated system
with respect to $k_{y}$ is lifted in the presence of modulation with the
explicit presence of $k_{y}$ in $x_{0}.$ The $n=0$ landau level is different
from the rest as the energy of this level is zero and electrons in this level
do not contribute to diffusive conductivity calculated in the next section.
The rest of the Landau levels broaden into bands. The Landau bandwidths $\sim
G_{n}$ oscillates as a function of $n$ since $L_{n}(u)$ are oscillatory
functions of the index $n$.

Since we are interested in electron transport in the presence of both electric
and magnetic modulations, we consider an additional weak electric modulation
potential given as $V(x)=V_{0}\cos(Kx)$ on the system. Here $V_{0}$ is the
amplitude of modulation. We can determine the energy eigenvalues in the
presence of weak electric modulation where we take $V_{o}$ to be an order of
magnitude smaller than the Fermi Energy $\varepsilon_{F}=v_{F}\hslash k_{F}%
\ $with $k_{F}=\sqrt{2\pi n_{e}}$ is the magnitude of Fermi wave vector with
$n_{e}$ being the electron concentration. Hence we can apply standard first
order perturbation theory to determine the energy eigenvalues in the presence
of modulation. The first order correction in the energy eigenvalues when
electric modulation is present is given as%
\begin{equation}
\varepsilon(n,x_{0})=\varepsilon(n)+V_{0}F_{n}\cos(Kx_{0}) \label{4}%
\end{equation}
Here, $\ F_{n}=\frac{1}{2}\exp(-\frac{u}{2})[L_{n}(u)+L_{n-1}(u)]$,
$u=\frac{K^{2}l^{2}}{2}$ and, $L_{n}(u)$ and $L_{n-1}(u)\ $are Laguerre polynomials.

The energy eigenvalues of Dirac electrons in the presence of both modulations
can be expressed as%
\begin{equation}
\varepsilon(n,k_{y})=\hbar\omega_{g}\sqrt{n}+\omega_{0}\cos(Kx_{0})G_{n}%
+V_{0}F_{n}\cos(Kx_{0}). \label{5}%
\end{equation}
To better appreciate the modulation effects on the Landau levels we determine
the asymptotic expression for the bandwidth ($\Delta$) next. The width of the
$n$th Landau level in the presence of periodic electric and magnetic
modulation is given as $\Delta=\Delta_{B}+\Delta_{E},$ where $\Delta_{E}$\ is
width of the electric modulation and $\Delta_{B}$\ is the width of the
magnetic modulation:%

\begin{equation}
\Delta_{B}=2\left\vert G_{n}\right\vert =2\sqrt{\frac{2n}{u}}e^{-u/2}%
[L_{n-1}(u)-L_{n}(u)] \label{6}%
\end{equation}
The asymptotic expression of bandwidth can be obtained by using the following
asymptotic expression for the Laguerre polynomials by taking the large $n$
($n_{F}=\frac{\varepsilon_{F}^{2}}{\varepsilon^{2}(n)}=\frac{\hbar K_{F}^{2}%
}{2eB}$) limit as%
\[
exp^{-u/2}L_{n}(u)\rightarrow\frac{1}{\sqrt{\pi\sqrt{nu}}}\cos(2\sqrt
{nu}-\frac{\pi}{4})
\]
with the result that the asymptotic expression for $\Delta_{B}$ is%
\begin{equation}
\Delta_{B}=\frac{8\omega_{0}}{Kl}\sqrt{\frac{\hbar K_{F}^{2}}{2eB\pi^{2}R_{g}%
}}\sin\left(  \frac{2eB\pi R_{g}}{2a\hbar K_{F}^{2}}\right)  \sin\left(
\frac{2\pi R_{g}}{a}-\frac{\pi}{4}\right)  . \label{7}%
\end{equation}
Similarly, for electric modulation, the bandwidth $\Delta_{E}$ is given as%
\[
\Delta_{E}=2\left\vert F_{N}\right\vert =V_{0}\exp^{-\frac{u}{2}}\left\vert
L_{n}(u)+L_{n-1}(u)\right\vert \,.
\]
and the asymptotic expression for $\Delta_{E}$ is%
\begin{equation}
\Delta_{E}=V_{0}\left(  \frac{a}{\pi^{2}R_{g}}\right)  ^{\frac{1}{2}}%
\cos\left(  \frac{2eB\pi R_{g}}{2a\hbar K_{F}^{2}}\right)  \cos\left(
\frac{2\pi R_{g}}{a}-\frac{\pi}{4}\right)  . \label{8}%
\end{equation}
Therefore the bandwidth in the presence of both electric and magnetic
modulations can be expressed as%
\begin{align}
\Delta &  =\frac{8\omega_{0}}{Kl}\sqrt{\frac{\hbar K_{F}^{2}}{2eB\pi^{2}R_{g}%
}}\sin\left(  \frac{2eB\pi R_{g}}{2a\hbar K_{F}^{2}}\right)  \sin\left(
\frac{2\pi R_{g}}{a}-\frac{\pi}{4}\right)  +\nonumber\\
&  V_{0}\left(  \frac{a}{\pi^{2}R_{g}}\right)  ^{\frac{1}{2}}\cos\left(
\frac{2eB\pi R_{g}}{2a\hbar K_{F}^{2}}\right)  \cos\left(  \frac{2\pi R_{g}%
}{a}-\frac{\pi}{4}\right)  . \label{9}%
\end{align}
In Fig.(1) we present the bandwidths as a function of the magnetic field for
temperature $T=6K$, electron density $n_{e}=3\times10^{11}cm^{-2},$ the period
of modulation $a=350nm.$The strength of the electric modulation $V_{0}=0.2meV$
where as $B_{0}=0.004T$ which corresponds to $\omega_{0}=0.2meV$ with the
result that both the modulations have equal strengths. In the same figure we
have also shown the bandwidths when either the magnetic or electric modulation
alone is present. We observe that electric and magnetic bandwidths are out of
phase while the positions of the extrema of combined bandwidth are shifted
with respect to electric and magnetic bandwidths. The combined bandwidth when
both modulations are present will affect the conductivity and that is
considered in the next section.

\section{Magnetoconductivity with Periodic Electric and Magnetic Modulation:
In-phase}

To calculate the electrical conductivity in the presence of weak electric and
magnetic modulations we use Kubo formula to calculate the linear response to
applied external fields. In a magnetic field, the main contribution to Weiss
oscillations comes from the scattering induced migration of the Larmor circle
center. This is diffusive conductivity and we shall determine it following the
approach in \cite{7,10,11,12} where it was shown that the diagonal component
of conductivity $\sigma_{yy}$ can be calculated by the following expression in
the case of quasielastic scattering of electrons%
\begin{equation}
\sigma_{yy}=\frac{\beta e^{2}}{L_{x}L_{y}}\underset{\zeta}{%
{\displaystyle\sum}
}f(E_{\zeta})[1-f(E_{\zeta})]\tau(E_{\zeta})(\upsilon_{y}^{\zeta})^{2}
\label{10}%
\end{equation}
$L_{x}$, $L_{y}$, are the dimensions of the layer, $\beta=\frac{1}{k_{B}T}%
$\ is the inverse temperature with $k_{B}$ the Boltzmann constant, $f(E)$ is
the Fermi Dirac distribution function, $\tau(E)$\ is the electron relaxation
time and $\zeta$ denotes the quantum numbers of the electron eigenstate. The
diagonal component of the conductivity $\sigma_{yy}$ is due to modulation
induced broadening of Landau bands and hence it carries the effects of
modulation in which we are primarily interested in this work. $\sigma_{xx}$
does not contribute as the component of velocity in the $x$-direction is zero
here. The collisional contribution due to impurities is not taken into account
in this work.

The summation in Eq.(10) over the quantum numbers $\zeta$ can be written as%
\begin{equation}
\frac{1}{L_{x}L_{y}}\underset{\zeta}{%
{\displaystyle\sum}
}=\frac{1}{2\pi L_{x}}%
{\displaystyle\int\limits_{0}^{\frac{L_{x}}{l^{2}}}}
dk_{y}\underset{n=0}{\overset{\infty}{%
{\displaystyle\sum}
}}=\frac{1}{2\pi l^{2}}\underset{n=0}{\overset{\infty}{%
{\displaystyle\sum}
}} \label{11}%
\end{equation}
The component of velocity required in Eq.(10) can be calculated from the
following expression%
\begin{equation}
\upsilon_{y}^{\zeta}=-\frac{\partial}{\hbar\partial k_{y}}\varepsilon
(n,k_{y}). \label{12}%
\end{equation}
Substituting the expression for $\varepsilon(n,k_{y})$ obtained in Eq.(5) into
Eq.(12) yields
\begin{equation}
\upsilon_{y}^{\zeta}=\left[  \frac{2\omega_{0}u}{\hbar K}\sin(Kx_{0}%
)G_{n}(u)+\frac{2V_{0}u}{\hbar K}\sin(Kx_{0})F_{n}(u)\right]  \label{13}%
\end{equation}
As a result $\upsilon_{y}^{\zeta},$ the corresponding velocity given by
Eq.(13) contains the contribution from both the modulations (electric and
magnetic) obtained in Eq.(5) compared to one term in velocity component for
each\cite{10,11}. This term ($\upsilon_{y}^{\zeta}$) has important
consequences for the quantum transport phenomena in modulated systems.

With the results obtained in Eqs.(11), (12) and (13) we can express the
diffusive contribution to the conductivity given by Eq.(10) as%
\begin{equation}
\sigma_{yy}=A_{0}\phi\label{14}%
\end{equation}
where%
\begin{equation}
A_{0}=\frac{e^{2}\tau\beta}{\pi\hbar^{2}} \label{15}%
\end{equation}
and $\phi$ is given as
\begin{equation}
\phi=%
{\displaystyle\sum_{n=0}^{\infty}}
\frac{g(\varepsilon_{n})}{[g(\varepsilon_{n})+1)]^{2}}\left[  \frac{\sqrt
{u}e^{-u/2}V_{0}}{2}(L_{n}(u)+L_{n-1}(u))+\omega_{0}e^{-u/2}\sqrt{2n}%
(L_{n-1}(u)-L_{n}(u))\right]  ^{2} \label{16}%
\end{equation}
where $g(\varepsilon)=\exp[\beta(\varepsilon-\varepsilon_{F}]$ and
$\varepsilon_{F}$ is the Fermi energy.

In Fig.(2) we show the in-phase conductivity (the magnetic and the electric
modulations are in-phase) $\sigma_{yy}$ given by Eqs(14,15,16) as a function
of the inverse magnetic field for temperature $T=6K$, electron density
$n_{e}=3\times10^{11}cm^{-2},$ the period of modulation $a=350nm.$ The
dimensionless magnetic field $\frac{B^{\prime}}{B}$ is introduced where
$B^{\prime}=\frac{\hslash}{ea^{2}}=0.0054T$ for $a=350nm.$ The strength of the
electric modulation $V_{0}=0.2meV$ where as $B_{0}=0.004T$ which corresponds
to $\omega_{0}=0.2meV$ with the result that both the modulations have equal
strengths. In the same figure we have also shown the conductivity when either
the magnetic or electric modulation alone is present. The $\frac{\pi}{2}$
phase difference in the bandwidths results in the same phase difference
appearing in the conductivity for electric and magnetic modulations as can be
seen in the figure. To better understand the effects of in-phase modulations
on the conductivity we consider the asymptotic expression of the quantity
$\phi$ given by Eq.(16) that appears in the magnetoconductivity $\sigma_{yy}$.
The asymptotic results are valid when applied magnetic field is weak such that
many Landau levels are filled. The asymptotic expression is obtained in the
next section.

\section{Asymptotic Expressions: In-Phase Modulations}

To get a better understanding of the results of the previous section we will
consider the asymptotic expression of conductivity where analytic results in
terms of elementary functions can be obtained following \cite{7,10,11}. The
asymptotic expression of $\phi$ can be obtained by employing the following
asymptotic expression for the Laguerre polynomials which is valid in the limit
of large $n$ when many landau levels are filled \cite{13}%
\begin{equation}
\exp^{-u/2}L_{n}(u)\rightarrow\frac{1}{\sqrt{\pi\sqrt{nu}}}\cos(2\sqrt
{nu}-\frac{\pi}{4}) \label{17}%
\end{equation}
with the result that the in-phase bandwidth can be written as%
\[
\Delta(\text{in-phase})=\frac{4\omega_{0}\times\sqrt{\frac{2n}{u}}\times
\sin\left(  1/2\sqrt{\frac{u}{n}}\right)  \times\sqrt{1+\delta^{2}}}{\sqrt
{\pi\sqrt{nu}}}\times\sin\left(  2\sqrt{nu}-\frac{\pi}{4}+\Phi\right)
\]
where the ratio between the two modulation strengths $\delta=\frac{V_{0}%
\cos\left(  1/2\sqrt{\frac{u}{n}}\right)  }{2\omega_{0}\sqrt{\frac{2n}{u}}%
\sin\left(  1/2\sqrt{\frac{u}{n}}\right)  }=\tan(\Phi).$ The flat band
condition from the above equation is $2\sqrt{nu}-\frac{\pi}{4}+\Phi=i\pi$
where $i$ is an integer. This condition can also be expressed as $\frac
{\sqrt{2n}}{a}l=i+\frac{1}{4}-\frac{\Phi}{\pi}$, where $n=n_{F}=\frac
{\varepsilon_{F}^{2}}{\varepsilon^{2}(n)}=\frac{\varepsilon_{F}^{2}}%
{\hslash^{2}\omega_{g}^{2}}$ is the highest Fermi integer. We see that the
flat band condition in this case depends on the relative strength of the two
modulations. We \ now take the continuum limit:%
\begin{equation}
n-->\frac{1}{2}\left(  \frac{l\varepsilon}{v\hbar}\right)  ^{2},\overset
{\infty}{\underset{n=0}{%
{\displaystyle\sum}
}}-->\left(  \frac{l}{v\hbar}\right)  ^{2}%
{\displaystyle\int\limits_{0}^{\infty}}
\varepsilon d\varepsilon\label{18}%
\end{equation}
to express $\phi$ in Eq.(16) as the following integral%
\begin{align}
\phi &  =\frac{4\omega_{0}^{2}\times\frac{2\varepsilon_{F}^{2}}{u\hslash
^{2}\omega_{g}^{2}}\times\sin^{2}\left(  1/2\sqrt{\frac{u\hslash^{2}\omega
_{g}^{2}}{\varepsilon_{F}^{2}}}\right)  \times(1+\delta^{2})}{\pi\sqrt{u}%
}\times\left(  \frac{l}{v\hbar}\right)  ^{2}\times\nonumber\\
&
{\displaystyle\int\limits_{0}^{\infty}}
\frac{\varepsilon}{\sqrt{n}}d\varepsilon\frac{g(\varepsilon)}{[g(\varepsilon
)+1)]^{2}}\sin^{2}(1/2\sqrt{u/n})\sin^{2}(2\sqrt{nu}-\frac{\pi}{4}+\Phi)
\label{19}%
\end{align}
where $u=2\pi^{2}/b$ with $b=\frac{eBa^{2}}{\hbar}=\frac{B}{B^{\prime}}$ and
$B^{\prime}=\frac{ea^{2}}{\hslash}$.

Now assuming that the temperature is low such that $\beta^{-1}\ll
\varepsilon_{F}$ and replacing $\varepsilon=\varepsilon_{F}+s\beta^{-1}$, we
rewrite the above integral as%
\begin{align}
\phi &  =\frac{4\omega_{0}^{2}\times\frac{2\varepsilon_{F}^{2}}{u\hslash
^{2}\omega_{g}^{2}}\times\sin^{2}\left(  1/2\sqrt{\frac{u\hslash^{2}\omega
_{g}^{2}}{\varepsilon_{F}^{2}}}\right)  \times(1+\delta^{2})}{\pi\sqrt{u}%
\beta}\times\left(  \frac{l}{v\hbar}\right)  ^{2}\times\nonumber\\
&
{\displaystyle\int\limits_{-\infty}^{\infty}}
\frac{dse^{s}}{(e^{s}+1)^{2}}\sin^{2}(\frac{2\pi p}{b}-\frac{\pi}{4}%
+\Phi+\frac{\sqrt{2u}}{v\sqrt{B\hbar}\beta}s) \label{20}%
\end{align}
where $p=\frac{\varepsilon_{F}a}{\hbar v}=k_{F}a=\sqrt{2\pi n_{s}}a$ is the
dimensionless Fermi momentum of the electron. To obtain an analytic solution
we have also replaced $\varepsilon$ by $\varepsilon_{F}$ in the above integral
except in the sine term in the integrand.

The above expression can be written as
\begin{align}
\phi &  =\frac{4\omega_{0}^{2}\times\frac{2\varepsilon_{F}^{2}}{u\hslash
^{2}\omega_{g}^{2}}\times\sin^{2}\left(  1/2\sqrt{\frac{u\hslash^{2}\omega
_{g}^{2}}{\varepsilon_{F}^{2}}}\right)  \times(1+\delta^{2})}{\pi\sqrt{u}%
\beta}\times\left(  \frac{l}{v\hbar}\right)  ^{2}\times\nonumber\\
&
{\displaystyle\int\limits_{-\infty}^{\infty}}
\frac{ds}{\cosh^{2}(s/2)}\sin^{2}(\frac{2\pi p}{b}-\frac{\pi}{4}+\Phi
+\frac{2\pi a}{vb\beta}s) \label{21}%
\end{align}
The above integration can be performed by using the following identity
\cite{13}:%
\begin{equation}%
{\displaystyle\int\limits_{0}^{\infty}}
dx\frac{\cos ax}{\cosh^{2}\beta x}=\frac{a\pi}{2\beta^{2}\sinh(a\pi/2\beta)}
\label{22}%
\end{equation}
with the result%
\begin{align}
\phi &  =\frac{2\omega_{0}^{2}\times\frac{2\varepsilon_{F}^{2}}{u\hslash
^{2}\omega_{g}^{2}}}{4\pi^{2}}\times\frac{T}{T_{D}}\times\sin^{2}\left(
\frac{\pi}{p}\right)  \times(1+\delta^{2})\times\nonumber\\
&  \left[  1-A\left(  \frac{T}{T_{D}}\right)  +2A\left(  \frac{T}{T_{D}%
}\right)  \sin^{2}\left[  2\pi\left(  \frac{p}{b}-\frac{1}{8}+\frac{\Phi}%
{2\pi}\right)  \right]  \right]  \label{23}%
\end{align}
where $k_{B}T_{D}=\frac{\hbar vb}{4\pi^{2}a},$ $\frac{T}{T_{D}}=\frac{4\pi
^{2}a}{\hbar vb\beta}$and $A(x)=\frac{x}{\sinh(x)}-^{(x-->\infty)}%
->=2xe^{-x}.$

From the asymptotic expression of $\phi$ given by Eq.(23), we observe that the
effect of the in-phase electric and magnetic modulations is the appearance of
a phase factor $\Phi$ in the conductivity. The shift in the Weiss oscillations
when in-phase electric and magnetic modulations are present can be seen in
Fig.(3). The phase factor $\Phi$ depends on the relative strength of the two
modulations. How the Weiss oscillations are affected as $\Phi$ as well as the
magnetic field is varied can be seen in Fig.(3). The results shown are for a
fixed magnetic modulation of strength $\omega_{0}=0.2meV$ and the electric
modulation is varied. The change in $V_{0}$ results in a corresponding change
in both $\delta$ and $\Phi$. From Fig.(3), we observe that the position of the
extrema in magnetoconductivity as a function of the inverse magnetic field
depends on the relative strength of the modulations.

The effects of electric and magnetic modulations that are out-of-phase on the
conductivity can be better appreciated if we consider the asymptotic
expression for $\phi$ in this case. This is taken up in the next section.

\section{Magnetoconductivity with Periodic Electric and Magnetic Modulation:
Out-of-phase}

In this section, we calculate $\phi$ when electric and magnetic modulations
are out of phase by $\pi/2.$ We consider magnetic modulation out of phase with
the electric one: We take the electric modulation to have the same phase as
given in the previous section with the $\pi/2$ phase difference incorporated
in the magnetic field. The energy eigenvalues are%
\begin{equation}
\varepsilon(n,k_{y})=\hbar\omega_{g}\sqrt{n}+\omega_{0}\sin(Kx_{0})G_{n}%
+V_{0}F_{n}\cos(Kx_{o}), \label{24}%
\end{equation}
and the bandwidth is%
\begin{equation}
\Delta(\text{out of phase})=\frac{4\omega_{0}\times\sqrt{\frac{2n}{u}}%
\times\sin\left(  1/2\sqrt{\frac{u}{n}}\right)  }{\sqrt{\pi\sqrt{nu}}}%
\times\sqrt{\delta^{2}+(1-\delta^{2})\sin\left(  2\sqrt{nu}-\frac{\pi}%
{4}\right)  }\,. \label{25}%
\end{equation}
The term responsible for Weiss oscillations is the $\sin\left(  2\sqrt
{nu}-\frac{\pi}{4}\right)  $ term under the square root which can be readily
seen by considering the large $n$ limit of the bandwidth. Therefore for
$\delta=\pm1$ Weiss oscillations are no longer present in the bandwidth. The
velocity component $\upsilon_{y}$ is given as%
\begin{equation}
\upsilon_{y}=-\left[  \frac{2\omega_{0}u}{\hbar K}\cos(Kx_{0})G_{n}%
(u)-\frac{2V_{0}u}{\hbar K}\sin(Kx_{0})F_{n}(u)\right]  \label{26}%
\end{equation}
and $\phi$ is given as
\begin{equation}
\phi=%
{\displaystyle\sum_{n=0}^{\infty}}
\frac{g(\varepsilon_{n})}{[g(\varepsilon_{n})+1)]^{2}}\left[  \frac
{ue^{-u}V_{0}^{2}}{4}(L_{n}(u)+L_{n-1}(u))^{2}+\omega_{0}^{2}e^{-u}%
2n(L_{n-1}(u)-L_{n}(u))^{2}\right]  . \label{27}%
\end{equation}
The asymptotic expression for $\phi$ in the presence of both electric and
magnetic modulations that are out of phase is obtained by substituting the
asymptotic expressions for the Laguerre polynomials and converting the sum
into integration with the result
\begin{align}
\phi &  =\frac{2\omega_{0}^{2}p^{2}}{\pi^{2}}\sin^{2}\left(  \frac{\pi}%
{p}\right)  \frac{T}{4\pi^{2}T_{D}}\times\nonumber\\
&  \left[  2\delta^{2}+(1-\delta^{2})\left(  1-A\left(  \frac{T}{T_{D}%
}\right)  +2A\left(  \frac{T}{T_{D}}\right)  \sin^{2}\left[  2\pi\left(
\frac{p}{b}-\frac{1}{8}\right)  \right]  \right)  \right]  . \label{28}%
\end{align}
From the expression of the out-of-phase bandwidth given by Eq.(25) we find
that Weiss oscillations in the bandwidth are absent for relative modulation
strength $\delta=\pm1$, the same is reflected in $\phi$ as the term
responsible for Weiss oscillations ($\sin^{2}\left[  2\pi\left(  \frac{p}%
{b}-\frac{1}{8}\right)  \right]  $) vanishes for $\delta=\pm1$ as can be seen
from the above equation. Therefore the magnetoconductivity $\sigma_{yy}$ does
not exhibit Weiss oscillations when the relative modulation strength
$\delta=\pm1.$The conductivity as a function of magnetic field when the
electric and magnetic modulations are out-of-phase is shown in Fig.(4). The
results shown are for a fixed magnetic modulation of strength $\omega
_{0}=0.2meV$ and the electric modulation $V_{0}$ is allowed to vary between
positive and negative values. The other parameters are the same as in
Figs.(1,2,3). As $V_{0}$ is varied there is a corresponding change in
$\delta.$ We find that the positions of the extrema of $\sigma_{yy}$ as a
function of the inverse magnetic field do not change as $\delta$ is varied
since the phase factor $\Phi$ does not appear in the expression of
conductivity when the two modulations are out of phase. It is also observed in
Fig.(4) that there is a $\frac{\pi}{2}$ phase difference between the curves
for $\delta\geq1$ and $\delta<1.$ The same behavior is observed in the
bandwidth which is reflected in the magnetoconductivity. In this work we have
considered Weiss oscillations and have not taken pure Shubnikov de Hass (SdH)
oscillations into account but SdH oscillations do appear superimposed on Weiss
oscillations in the region of strong magnetic field as can been seen in all of
our figures. In addition, since this work was motivated by\cite{7}, it is
important to compare the results obtained here for graphene monolayer with
those presented for standard 2DEG. In contrast to the work presented here, in
\cite{7} SdH are explicitly taken into account but we find that our principal
expressions Eq.(23), (28) reduce to the expressions there when the terms
contributing to SdH oscillations are ignored.

In the end, we would also like to mention relevance of this work in
experimental studies of transport in graphene. Experiments on graphene
monolayer in the presence of modulated electric and magnetic fields have not
been realized yet, this theoretical work is in anticipation of experimental
work. We expect that modulation effects predicted here can be observed in
graphene employing established techniques that were used for the
two-dimensional electron gas systems found in semiconductor heterostructures
\cite{14}. In conventional semiconductor systems, modulation of the potential
seen by electrons can be produced by molecular beam epitaxy, chemical vapor
deposition as well as sputtering techniques \cite{15}. In graphene, we expect
that modulation effects can be introduced by adsorbing adatoms on graphene
surface using similar techniques, by positioning and aligning impurities with
scanning tunneling microscopy or by applying top gates to graphene. Epitaxial
growth of graphene on a patterned substrate is also possible \cite{16}.

In conclusion, we have determined the effects of both the electric and
magnetic modulations on the magnetoconductivity of a graphene monolayer.
Appearance of Weiss oscillations in the magnetoconductivity $\sigma_{yy}$ is
the main focus of this work. These oscillations are affected by the relative
phase of the two modulations and position of the oscillations depends on the
relative strength of the two modulations. We find complete suppression of
Weiss oscillations for particular relative strength of the modulations when
the modulations are out-of-phase.

\section{Acknowledgements}

We gratefully acknowledge F. M. Peeters for suggesting the problem and for
comments on the manuscript. One of us (K.S.) would like to acknowledge the
support of the Pakistan Science Foundation (PSF) through project No. C-QU/Phys
(129). M. T. would like to acknowledge the support of the Pakistan Higher
Education Commission (HEC).

$\ast$Department of Physics, Blackett Laboratory, Imperial College London,
South Kensington Campus, London SW7 2AZ, United Kingdom.; Electronic address: m.tahir06@imperial.ac.uk

$\dagger$Electronic address: ksabeeh@qau.edu.pk; kashifsabeeh@hotmail.com


\begin{thebibliography}{99}                                                                                               %


\bibitem {1}K. S. Novoselov\textit{, }A. K. Geim, S. V. Morozov, D. Jiang, M.
I. Katsnelson, I. V. Grigorieva, S. V. Dubonos, and A. A. Firsov, Nature
\textbf{438,} 197 (2005); Y. Zhang, Y. W. Tan, H. L. Stormer, and P. Kim,
\textbf{438}, 201 (2005).

\bibitem {2}Y. Zheng and T. Ando, Phys. Rev. B \textbf{65}, 245420 (2002); V.
P. Gusynin and S. G. Sharapov, Phys. Rev. Lett. \textbf{95}, 146801 (2005); N.
M. R. Perez F. Guinea, and A. H. Castro Neto, Phys. Rev. B \textbf{73}, 125411
(2006); M. I. Katsnelson , K. S. Novoselov, and A. K. Geim, Nat. Phys.
\textbf{2}, 620 (2006); K. S. Novoselov, E. McCann, S. V. Morozov, V. I.
Fal'ko, M. I. Katsnelson, U. Zeitler, D. Jiang, F. Schedin, and A. K. Geim,
Nat. Phys. \textbf{2}, 177 (2006).

\bibitem {3}D. Weiss, K. v. Klitzing, K. Ploog, and G. Weimann\textit{,
}Europhys. Lett., \textbf{8}, 179 (1989)

\bibitem {4}R. W. Winkler, J. P. Kotthaus, and K. Ploog,\textit{ }Phys. Rev.
Lett. \textbf{62}, 1177 (1989).

\bibitem {5}R. R. Gerhardts, D. Weiss, and K. v. Klitzing,\textit{ }Phys. Rev.
Lett. \textbf{62}, 1173 (1989).

\bibitem {6}C. Berger,\textit{et.al}, Science \textbf{312}, 1191 (2006).

\bibitem {7}F. M. Peeters and P. Vasilopoulos, Phys. Rev. B \textbf{47}, 1466 (1993)

\bibitem {8}F. M. Peeters and P. Vasilopoulos, Phys. Rev. B \textbf{46}, 4667
(1992); A. Manolescu, R. R. Gerhardts, Phys. Rev. B \textbf{56}, 9707 (1997);
R. R. Gerhardts, Phys. Rev. B \textbf{53}, 11064 (1996); R. Menne, R. R.
Gerhardts, Phys. Rev. B \textbf{57}, 1707 (1998); U. J. Gossmann, A.
Manolescu, R. R. Gerhardts, Phys. Rev. B \textbf{57}, 1680 (1998); J. Shi and
F. M. Peeters, K. W. Edmonds and B. L. Gallagher, Phys. Rev. B \textbf{66},
035328 (2002); P. De Ye, D. Weiss, R. R. Gerhardts, M. Seeger, K. v.
Klitzing,\textit{ }K. Eberl and H. Nickel, Phys. Rev. Lett. \textbf{74}, 3013
(1995); J. H. Ho, Y. H. Lai, Y. H. Chui and M. F. Lin, Nanotechnology
\textbf{19}, 035712 (2008);

\bibitem {9}P. Vasilopoulos, F. M. Peeters, Superlattices and Microstructures
\textbf{7}, 393 (1990); F. M. Peeters and A. Matulis, Phys. Rev. B
\textbf{48}, 15166 (1993); F. M. Peeters, P. Vasilopoulos and Jirong Shi, J.
Phys. : Condens. Matter \textbf{14}, 8803 (2002); X. F. Wang and P.
Vasilopoulos, F. M. Peeters, Phys. Rev. B \textbf{71}, 125301 (2005); X. F.
Wang, P. Vasilopoulos and F. M. Peeters, Phys. Rev. B \textbf{69}, 035331
(2004); D. P. Xue and G. Xiao, Phys. Rev. B \textbf{45}, 5986 (1992).

\bibitem {10}A. Matulis and F. M. Peeters, Phys. Rev. B \textbf{75}, 125429 (2007).

\bibitem {11}M. Tahir, K. Sabeeh, Phys. Rev. B \textbf{77}, 195421 (2008).

\bibitem {12}M. Charbonneau\textit{, }K. M. Van Vliet and P.
Vasilopoulos,\textit{ }J. Math. Phys. \textbf{23}, 318 (1982).

\bibitem {13}I. S. Gradshteyn and I. M. Ryzhik, \textit{Table of Integrals,
Series and Products} (Academic Press, New York, 1980).

\bibitem {14}H. A. Carmona,\textit{et.al}, Phys. Rev. Lett. \textbf{74}, 3009
(1995); K. W. Edmonds,\textit{et.al}, Phys. Rev. B \textbf{64}, 041303 (2001);
P. De Ye, D. Weiss, R. R. Gerhardts, M. Seeger, K. v. Klitzing,\textit{ }K.
Eberl and H. Nickel, Phys. Rev. Lett. \textbf{74}, 3013 (1995).

\bibitem {15}R. Tsu, \textit{Superlattice to Nanoelectronics} (Elsevier,
Oxford, UK, 2005); M. G. Cottam, D. R. Tilley, \textit{Introduction to Surface
and Superlattice Excitations} (Cambridge Univ. Press, UK, 1989).

\bibitem {16}C- H Park,\textit{et.al}, Nat. Phys. \textbf{4, }213 (2008).
\end{thebibliography}
\end{document}